\documentclass[11pt]{article}
\usepackage{amssymb}
\newcommand{\equal}{\!\!\!&=&\!\!\!}
\newcommand{\newequiv}{\!\!\!&\equiv&\!\!\!}
\begin{document}
\abovedisplayshortskip 12pt
\belowdisplayshortskip 12pt
\abovedisplayskip 12pt
\belowdisplayskip 12pt
\baselineskip=16pt
\title{{\bf The Short Pulse Hierarchy}} 
\author{J. C. Brunelli  \\
\\
Departamento de F\'\i sica, CFM\\
Universidade Federal de Santa Catarina\\
Campus Universit\'{a}rio, Trindade, C.P. 476\\
CEP 88040-900\\
Florian\'{o}polis, SC, Brazil\\
}
\date{}
\maketitle

\begin{center}
{ \bf Abstract}
\end{center}

We study a new hierarchy of equations containing the Short Pulse equation, which describes the evolution of very short pulses in nonlinear media, and the Elastic Beam equation, which describes nonlinear transverse oscillations of elastic beams under tension. We show that the hierarchy of equations is integrable. We obtain the two compatible Hamiltonian structures. We construct an infinite series of both local and nonlocal conserved charges. A Lax description is presented for both systems. For the Elastic Beam equations we also obtain a nonstandard Lax representation.

\newpage

\section{Introduction:}

The cubic nonlinear Schr\"odinger equation is used in the description of the propagation of pulses in nonlinear media such as optical fibers. Recently, technology progress for creating very short pulses was achieved, however, the description of the evolution of those pulses lies beyond the usual approximations leading to the nonlinear Schr\"odinger equation. Various approaches have been proposed to replace the nonlinear Schr\"odinger equation in these conditions. In Ref. \cite{schafer} Sch\"afer and Wayne proposed an alternative model to approximate the evolution of very short pulses in nonlinear media. They derived the short pulse (SP) equation
\begin{equation}
u_{xt}=u+{1\over 6}\,\left( u^3\right)_{xx}\;.\label{xtshortpulse}
\end{equation}
Chung et al \cite{chung} showed numerically that the SP equation provides a better approximation to the solution of Maxwell's equations than the nonlinear Schr\"odinger equation as the pulse length gets short. Also, Sakovich and Sakovich \cite{sakovich} have studied the integrability of (\ref{xtshortpulse}).

In this paper we will study the integrability of the nonlocal representation of the SP equation (\ref{xtshortpulse}) 
\begin{equation}
u_{t}=(\partial^{-1}u)+{1\over 2}\,u^2u_x\;,\label{shortpulse}
\end{equation}
as well the hierarchy of equations associated with it. The equation in this form is more feasible for a Hamiltonian description and can be obtained integrating (\ref{xtshortpulse}) with respect to $x$. In what follows we will refer to the equation (\ref{shortpulse}) simply as the SP equation.

The local nonlinear equation
\begin{equation}
u_{t}=\left[{u_{xx}\over(1+u_x^2)^{3/2}}\right]_x\;,\label{elasticbeam}
\end{equation}
will also appear in our hierarchy of equations. It can be embedded in the Wadati-Konno-Ichikawa (WKI) system \cite{wadati} and its $x$ derivative can be shown to describe  nonlinear  transverse oscillation of elastic beams under tension \cite{ichikawa}. Therefore, we will simply call (\ref{elasticbeam}) the equation for elastic beams (EB).

Our paper is organized as follows. In Sec. 2, we shown that our system is integrable through its bi-Hamiltonian nature. We give the two Hamiltonian structures associated with the Hamiltonian description of equations (\ref{shortpulse})  and (\ref{elasticbeam}). The method of prolongation is used to prove the Jacobi identity as well the compatibility of the Hamiltonian structures. In Sec. 3, we construct the recursion operator and its inverse to obtain the first local and nonlocal charges recursively. We also obtain the first local and nonlocal equations of the hierarchy  recursively, which includes the EB and SP equations, respectively.  In Se.c 4 we obtain the Lax representation for the system. For the EB equation, and for other local equations in the hierarchy, we also give a nonstandard Lax representation.  In Sec. 5, we summarize our results and  present our conclusions. 

The results in this paper are much like the ones obtained for the Harry Dym and Hunter-Saxton\footnote{In References \cite{brunelli3} and \cite{brunelli1} the equation obtained by Hunter and Saxton in {\it SIAM J. Appl. Math.} {\bf 51}, 1498 (1991) was erroneously named, by the present author, Hunter-Zheng. We sincerely apologize the authors about this mistake.} hierarchy of equations \cite{brunelli3}, deformed Harry Dym and Hunter-Saxton hierarchy of equations \cite{brunelli1} and for the Non-local Gas hierarchy of equations \cite{brunelli9}. In these works we have as a main characteristic a hierarchy of integrable equations with positive and negative flows. Also, throughout this paper the calculations involving pseudo-differential operators were performed or checked by the computer algebra program PSEUDO \cite{brunelli}.

\section{Bi-Hamiltonian Structure:}

Following \cite{brunelli1} let us introduce \begin{equation}                     
F^2\equiv 1+u_x^2\;,\quad A\equiv {u_x\over F}\;,\label{fa}  
\end{equation}
which satisfy the following useful properties
\begin{eqnarray}                                                                    
F^2(1\!\!\!\!&-&\!\!\!\!A^2)= 1\;,\nonumber\\\noalign{\vskip 5pt} 
\frac{F_{x}}{F^3} \equal AA_x\;,\nonumber\\\noalign{\vskip 5pt}          
{u_{xx}\over F^3}\equal A_x\;.\label{identities}            
\end{eqnarray}                                                                      
Using (\ref{fa}) the SP equation (\ref{shortpulse}) assumes the forms
\begin{eqnarray}
F_{t} \equal \left({1\over2}u^2F\right)_x\;,\nonumber\\\noalign{\vskip 5pt}
A_{t} \equal {u\over F}+{1\over2}u^2A_x\;,\label{fsp}
\end{eqnarray}
and, similarly, the EB equation (\ref{elasticbeam}) can be written in one of the forms
\begin{eqnarray}
u_t \equal A_{xx}\;,\nonumber\\\noalign{\vskip 5pt}
F_{t} \equal AA_{xxx}\;,\nonumber\\\noalign{\vskip 5pt}
A_{t} \equal {A_{xxx}\over F^3}\;.\label{fEB}
\end{eqnarray}                      
Let us stress that the basic field is $u$ and that $A$ and $F$ are just placeholders used to make expressions more compact. Also, it is interesting to observe that the EB equation when written in the form (\ref{fEB}) has the same structural form of the deformed Harry Dym equation studied in \cite{brunelli1}, however, the definitions in (\ref{fa}) are different.

From (\ref{fsp}) and (\ref{fEB}) it is straightforward to note that
\begin{equation}
H_{0}=-\int dx\,F\label{h0}\;,
\end{equation}
is conserved under both the SP and EB flows.

Introducing the Clebsch potential
\begin{equation}
u=\phi_x\;,\label{transf}
\end{equation}
the equation (\ref{shortpulse}) can be written as
\begin{equation}
\phi_t=(\partial^{-1}\phi)+{1\over 6}\phi_x^3\;.\label{phishortpulse}
\end{equation}
This equation (\ref{phishortpulse}) can be obtained from a variational principle,
$\delta\int dtdx\,{\cal L}$, with the Lagrangian density
\begin{equation}
{\cal L}={1\over 2}\phi_t \phi_x - {1\over24} \phi_x^4 + {1\over 2}\phi^2\;.\label{lagrangian}
\end{equation}
This is a first order Lagrangian density and, consequently, the
Hamiltonian structure can be readily read out, or we can use, for
example, Dirac's theory of constraints \cite{dirac} to obtain the
Hamiltonian and the Hamiltonian operator associated with
(\ref{lagrangian}). The Lagrangian is degenerate and the primary
constraint is obtained to be
\begin{equation}
\Phi=\pi-{1\over 2}\phi_x\;,\label{primary}
\end{equation}
where $\pi={{\partial{\cal L}}/{\partial \phi_t}}$ is the canonical
momentum. The total Hamiltonian can be written as
\begin{equation}
H_T =\int dx\left(\pi \phi_t-{\cal L}+\lambda\Phi\right)
=\int dx\left[ {1\over24}\phi_x^4 - {1\over 2}\phi^2 +\lambda\left(\pi-{1\over
2}\phi_x\right)\right]\;,\label{ht}
\end{equation}
where $\lambda$ is a Lagrange multiplier field. Using the canonical
Poisson bracket relation
\begin{equation}
\{\phi(x),\pi(y)\}=\delta(x-y)\;,\label{poisson}
\end{equation}
with all others vanishing, it follows that the requirement of the
primary constraint to be stationary under time evolution,
\[
\{\Phi(x),H_T\}=0\;,
\]
determines the Lagrange multiplier field $\lambda$ in (\ref{ht}) and
the system has no further constraints.

Using the canonical Poisson bracket relations (\ref{poisson}), we
can now calculate
\begin{equation}
K(x,y)\equiv\{\Phi(x),\Phi(y)\}={1\over
2}\partial_y\delta(y-x)-{1\over
2}\partial_x\delta(x-y)\;.\label{kpoisson}
\end{equation}
This shows that the constraint (\ref{primary}) is second class and
that the Dirac bracket between the basic variables has the form
\[
\{\phi(x),\phi(y)\}_D=\{\phi(x),\phi(y)\}-\int
dz\,dz'\{\phi(x),\Phi(z)\}J(z,z')\{\Phi(z'),\phi(y)\}=J(x,y)\;,\
\]
where $J$ is the inverse of the Poisson bracket of the constraint
(\ref{kpoisson}),
\[
\int dz\,K(x,z) J(z,y)=\delta(x-y)\,.
\]
This last relation determines
\[
\partial_x J(x,y)=-\delta(x-y)\;,
\]
or
\[
J(x,y)={\cal D}\delta(x-y)\;,
\]
where
\begin{equation}
{\cal D}=-\partial^{-1}\;,\label{d}
\end{equation}
and can be thought of as the alternating step function in the
coordinate space. We can now set the constraint (\ref{primary})
strongly to zero in (\ref{ht}) to obtain
\begin{equation}
H_T =\int dx\left({1\over24}\phi_x^4-{1\over
2}\phi^2\right)\;.\label{h}
\end{equation}
Using (\ref{transf}) and the transformation properties of Hamiltonian operators \cite{blaszak}, we get
\begin{equation}
{\cal D}_1=\partial\left({\cal D}\right)(\partial)^*=\partial\;,\label{d1}
\end{equation}
and the SP equation (\ref{shortpulse}) can be written in the Hamiltonian form as
\[
u_t={\cal D}_1{\delta H_2\over\delta u}\;,
\]
with $H_{2}$ given by 
\begin{equation}
H_2 =\int dx\left[{1\over24}u^4-{1\over
2}(\partial^{-1}u)^2\right]\;,\label{h2}
\end{equation}
which can be easily checked to be conserved by both the SP and EB equations. 

We will show that the SP and EB equations belong to the same hierarchy of equations, at this point we note that
\[
u_t={\cal D}_1{\delta H_0\over\delta u}\;,
\]
with $H_0$ given by (\ref{h0}), yields the EB equation (\ref{elasticbeam}).

It is easy to show that the charges
\begin{eqnarray}
H_{-1} \equal {1\over2}\int dx\,F A_{x}^{2}\label{h-1}\;,\\\noalign{\vskip 5pt}
H_1 \equal{1\over2}\int dx \,u^2\label{h1}
\end{eqnarray}
are also conserved by both the SP and EB equations. Therefore, the SP equation (\ref{shortpulse}) can be written in the Hamiltonian form as
\[
u_t={\cal D}_2{\delta H_1\over\delta u}\;,
\]
and the EB equation (\ref{elasticbeam}) as
\[
u_t={\cal D}_2{\delta H_{-1}\over\delta u}\;,
\]
where we have defined
\begin{equation}
{\cal D}_2=\partial^{-1}+u_x\partial^{-1}u_x=\left(F^2-u_x\partial^{-1}u_{xx}\right)\partial^{-1}\;.\label{d2}
\end{equation}
The skew symmetry of this Hamiltonian
structures is manifest. The proof of the Jacobi identity for this
structure as well its compatibility with (\ref{d1}) can be shown through the standard method of
prolongation \cite{olver} which we describe briefly. 

We can construct the two bivectors associated with the two
structures as
\begin{eqnarray}
\Theta_{{\cal D}_1}\equal{1\over 2}\int dx\,\left\{\theta\wedge{\cal
D}_1\theta\right\}={1\over2}\int
dx\,\theta\wedge\theta_{x}\;,\nonumber\\\noalign{\vskip 5pt}
\Theta_{{\cal D}_2}\equal{1\over 2}\int dx\,\left\{\theta\wedge{\cal
D}_2\theta\right\}={1\over2}\int
dx\,\left\{\theta\wedge(\partial ^{-1}\theta)+u_x\,\theta\wedge(\partial^{-1}u_x\,\theta)\right\}\;.\nonumber
\end{eqnarray}
Using the prolongation relations,
\begin{eqnarray}
\hbox{\bf pr}\,{\vec v}_{{\cal D}_1{\theta}} (u)
\equal\theta_{x}\nonumber\;,\\
\noalign{\vskip 5pt} 
\hbox{\bf pr}\,{\vec v}_{{\cal D}_2{\theta}} (u) 
\equal (\partial^{-1}\theta)+u_x(\partial^{-1}u_x\,\theta)\nonumber\;,\\
\noalign{\vskip 5pt} 
\hbox{\bf pr}\,{\vec v}_{{\cal D}_2{\theta}} (u_x) \equal\left(\hbox{\bf pr}\,{\vec
v}_{{\cal D}_2{\theta}} (u)\right)_x\;,\nonumber\\
\label{prolongation}
\end{eqnarray}
it is straightforward to show that the prolongation of the
bivector $\Theta_{{\cal D}_2}$ vanishes,
\[
\hbox{\bf pr}\,{\vec v}_{{\cal D}_2\theta}\left(\Theta_{{\cal
D}_2}\right)=0\;,
\]
implying that ${\cal D}_2$ satisfies  Jacobi identity.  Using
(\ref{prolongation}), it also follows that
\[
\hbox{\bf pr}\,{\vec v}_{{\cal D}_1\theta}\left(\Theta_{{\cal
D}_2}\right)+\hbox{\bf pr}\,{\vec v}_{{\cal
D}_2\theta}\left(\Theta_{{\cal D}_1}\right)=0\;,
\]
showing that ${\cal D}_1$ and ${\cal D}_2$ are compatible. Namely,
not only are ${\cal D}_{1}, {\cal D}_{2}$ genuine Hamiltonian
structures, any arbitrary linear combination of them is as well.
As a result, the dynamical equations (\ref{shortpulse}) and (\ref{elasticbeam}) are bi-Hamiltonian with the same compatible Hamiltonian structures and, consequently, are integrable \cite{olver,magri}.

\section{The Short Pulse Hierarchy:}

When a system is  bi-Hamiltonian, we can naturally define a
hierarchy of commuting flows through the relation
\begin{equation}
u_{t_n}=K_n[u]={\cal D}_1{\delta H_{n+1}\over\delta u}={\cal
D}_2{\delta H_{n}\over\delta u}\;,\quad
n=\dots,-2,-1,0,1,2,\dots\;.\label{eqrecursion}
\end{equation}
For $n=1$ and $n=-1$ we get the SP and EB equations, respectively. For $n=0$ we have
\[
K_0={\cal D}_1{\delta H_{1}\over\delta u}=(\partial u)=u_x\;,
\]
and 
\begin{eqnarray}
K_0={\cal D}_2{\delta H_{0}\over\delta u}\equal\left(\left(F^2-u_x\partial^{-1}u_{xx}\right)\partial^{-1}A_x\right)\nonumber\\\noalign{\vskip 5pt}
\equal\left(F^2A-u_x\partial^{-1}u_{xx}A\right)=Fu_x-u_xF=0\;,\label{K0}
\end{eqnarray}
where we have used (\ref{identities}), i.e., $F^2A=Fu_x$ and $u_{xx}A=F_x$. Therefore, we would be lead to conclude that $H_0$ is a Casimir of ${\cal D}_2$. We can resolve this apparent contradiction being careful while performing calculations with the antiderivative $\partial^{-1}$. We can use the following representation
\[
(\partial_x^{-1} f)\equiv(\partial^{-1}f)(x)=\int_{-\infty}^{+\infty} dy\,\epsilon(x-y)f(y)\;, \label{dminus}
\]
where
\[
\epsilon(x-y)=\left\{\begin{array}{rl}
1/2&\hbox{ for }x>y\;,\\
-1/2&\hbox{ for } x<y\;.\end{array}\right.
\]
Then, it can be shown that
\begin{equation}
\left(\partial_x^{-1}f_x\right)=f-{1\over2}\left(f(+\infty)+f(-\infty)\right)\;,\label{dfx}
\end{equation}
Now, if we assume $u^{(n)}\to 0$ as $|x|\to\infty$ (which yields $A|_{\pm\infty}=0$ and $F|_{\pm\infty}=1$) then it follows 
\begin{equation}
(\partial^{-1}A_x)=A\quad\hbox{and}\quad(\partial^{-1}F_x)=F-1\;.\label{dAdF}
\end{equation}
Using this last result in the naive calculation (\ref{K0}) we obtain the desired term $u_x$. This sort of missing or ``ghost'' terms given rise to apparent contradictions in nonlocal theories were already observed in the literature (see \cite{ghost} and references therein).

Let us introduce the recursion operator following from the two
Hamiltonian structures as
\begin{equation}
R={\cal D}_2{\cal D}_1^{-1}\;.\label{r}
\end{equation}
Then, it follows from (\ref{eqrecursion}) that
\begin{equation}
{\delta H_{n+1}\over\delta u}=R^\dagger{\delta H_{n}\over\delta
u}\;,\quad n=0,1,2,\dots\;,\label{hrecursion}
\end{equation}
where
\begin{equation}
R^\dagger=\partial^{-2}+\partial^{-1}u_x\,\partial^{-1}u_x=
\partial^{-2}\left(F^2+u_{xx}\,\partial^{-1}u_x\right)\label{rdagger}
\end{equation}
is the adjoint of $R$. The conserved charges for the hierarchy
can, of course, be determined recursively from
(\ref{hrecursion}) and give the infinite set of (nonlocal) conserved Hamiltonians
\begin{eqnarray}
H_0\equal-\int dx\,F\;,\nonumber\\\noalign{\vskip 7pt}
H_1\equal{1\over2}\int dx\,u^2\;,\nonumber\\\noalign{\vskip 7pt}
H_2\equal\int dx\left[{1\over24}u^4-{1\over
2}(\partial^{-1}u)^2\right]\;,\nonumber\\\noalign{\vskip 7pt}
H_3\equal\int  dx\,\left[{1\over 720}u^6+{\frac{1}{2}}(\partial^{-2}u)^2+{\frac{1}{6}}(\partial^{-2}u^3)u-
{\frac{1}{4}}(\partial^{-1}u)^2u^2\right]\;,\nonumber\\\noalign{\vskip 7pt}
&\vdots&\;.\label{nonlocal}
\end{eqnarray}
The corresponding flows (the first few, since the equations become extremely nonlocal as we proceed further in the recursion) have the forms
\begin{eqnarray}
\begin{array}{l}
\displaystyle u_{t_{0}}=u_x\;,\\
\noalign{\vspace{15pt}}
\displaystyle u_{t_{1}}=(\partial^{-1}u)+{1\over 2}\,u^2u_x\;,\\
\noalign{\vspace{15pt}}
\displaystyle u_{t_{2}}=(\partial^{-3}u)+{1\over 6}(\partial^{-1}u^3)+ u_x\left(\partial^{-1}\left(u_x\left(\partial^{-2}u\right)\right)\right)+{1\over 24}\,u^4u_x\;,\\
\quad\,\,\,\vdots\;.
\end{array}\label{nonlocalflow}
\end{eqnarray}

For negative values of $n$, the gradients of the Hamiltonians will satisfy the recursion \cite{brunelli} 
\begin{equation}
{\delta H_{n}\over\delta u}=(R^\dagger)^{-1}{\delta H_{n+1}\over\delta
u}\;,\quad n=-1,-2,\dots\;.\label{inversehrecursion}
\end{equation}
Writing the recursion operator (\ref{r}) in the form
\begin{equation}
R=\left(F^2-u_x\,\partial^{-1}u_{xx}\right)\partial^{-2}\;,\label{rf}
\end{equation}
and using the identities (\ref{identities}) the inverse can be easily checked to be
\begin{equation}
R^{-1}=\partial^2\left({1\over F^2}+A\,\partial^{-1}{A_{x}}\right)=\partial^2{1\over{u_{xx}}}\partial F\partial^{-1}{u_{xx}\over F^3}\;.\label{invrf}
\end{equation}
Note that (\ref{invrf}) can be recognized as the recursion operator obtained in \cite{sakovich} for the SP equation in the form (\ref{xtshortpulse}) using cyclic basis techniques.
The corresponding conserved charges can now be recursively constructed from (\ref{inversehrecursion}) and have the forms 
\begin{eqnarray}
H_{-1} \equal {1\over2}\int dx\,F A_{x}^{2}\;,\nonumber\\\noalign{\vskip 7pt}
H_{-2}\equal {1\over8}\int  dx\,\left(FA_x^4-4{A_{xx}^2\over F}\right)\;,\nonumber\\\noalign{\vskip 7pt}
H_{-3}\equal{1\over 16}\int  dx\,\left(FA_x^6+8{AA_{xx}^3\over F}-12{A_x^2A_{xx}^2\over F}+8{A_{xxx}^2\over F^3}\right)\;,\nonumber\\ 
\noalign{\vskip 7pt}
&\vdots&\;.\label{local}
\end{eqnarray}
The corresponding flows (the first few) have the forms
\begin{eqnarray}
\begin{array}{l}
u_{t_{-1}}= A_{xx}\;,\\
\noalign{\vspace{10pt}}
u_{t_{-2}}=\displaystyle\left({ A_{xx}\over F^2}+{1\over2}A_x^2A\right)_{xx}\;,\\
\noalign{\vspace{10pt}}
u_{t_{3}}= \displaystyle\left({ A_{xxxx}\over F^4}+{1\over2}{A_x^2A_{xx}\over F^2}-{2}{A_{xxx}A_{x}A\over F^2}-{3\over2}{A_{xx}^2A\over F^2}-A_{xx}A_x^2A^2+{3\over8}A_x^4A\right)_{xx}\;,\\
\quad\,\,\,\vdots\;.
\end{array}\label{localflow}
\end{eqnarray}

\section{The Lax Representation:}

Conserved charges for our systems can be determined in principle recursively from
(\ref{hrecursion}) and (\ref{inversehrecursion}). However, to construct the conserved charges directly we look for a
Lax representation for the system of SP and EB equations.

It is well known \cite{okubo,brunelli2} that for a bi-Hamiltonian
system of evolution equations, $u_{t_n}=K_n[u]$, a natural Lax
description
\[
{\partial M\over\partial t_n}=[B,M]\;,
\]
is easily obtained where, we can identify
\begin{eqnarray}
M \newequiv R\;,\nonumber\\
\noalign{\vskip 5pt} B
\newequiv K'_n\;.\nonumber
\end{eqnarray}
Here $R$ is the recursion operator (\ref{rf}) and $K'_n$ represents the Fr\'echet derivative of $K_n$, defined
by
\[
K'_n[u]\,v=\frac{d\ }{d\epsilon}\,K_n[u+\epsilon
v]\Big|_{\epsilon=0}\Big.\;.
\]
For the SP and EB system of equations in (\ref{shortpulse}) and
(\ref{elasticbeam}) respectively, we have
\begin{eqnarray}
B^{\hbox{\scriptsize\,SP}}\newequiv
K'_{1}=\partial^{-1}+{1\over2}\,\partial u^2
\;,\nonumber\\\noalign{\vskip 5pt}
B^{\hbox{\scriptsize\,EB}}\newequiv
K'_{-1}=\partial^2 F^{-3}\partial
\;.\nonumber
\end{eqnarray}
The two systems have the same $M=R$ given in
(\ref{r}). It can now be checked that
\begin{eqnarray}
\frac{\partial M}{\partial t} \equal \left[B^{\hbox{\scriptsize
\,SP}},M\right]\;,\nonumber\\\noalign{\vskip 5pt}
\frac{\partial M}{\partial t} \equal \left[B^{\hbox{\scriptsize
\,EB}},M\right]\;,\label{lax}
\end{eqnarray}
do indeed generate the SP and the EB equations and, thereby,
provide a Lax pair for the system.

A Lax representation directly gives the conserved charges of the system. From the
structure of (\ref{lax}), it follows that ${\rm
Tr}\,M^{\frac{2n+1}{2}}$ are conserved, where ``Tr" represents the
Adler's trace \cite{adler}. We note that
\begin{eqnarray}
\hbox{Tr}M^{2n+1\over 2} \equal 0\;,\quad
n\ge1\nonumber\\\noalign{\vskip 5pt} \hbox{Tr}M^{{1\over 2}} \equal
\int dx\, F\;,\nonumber\\\noalign{\vskip 5pt}
\hbox{Tr}M^{-{1\over2}} \equal {1\over2}\int dx\,F
A_{x}^{2}\;,\nonumber\\\noalign{\vskip 5pt}
\hbox{Tr}M^{-{3\over2}} \equal {3\over8} \int dx\,\left( F
A_{x}^{4}-4\,
\frac{A_{xx}^{2}}{F}\right)\;,\nonumber\\
&\vdots&\;.\label{charges}
\end{eqnarray}
These charges correspond (up to multiplicative constants) to the ones given in (\ref{local}),
constructed earlier by recursion. In fact, all $H_{-n}$ with
positive $n\geq 0$ can be constructed from ${\rm
Tr}\,M^{-\frac{2n-1}{2}}$ and by construction (namely, because of
the nature of (\ref{lax})), they are conserved under both the SP
and EB flows. However, as is clear from (\ref{charges}),
this procedure does not yield the nonlocal charges $H_{n}$  with positive
integer values. The
 construction of these nonlocal charges relies primarily on
the recursion relation (\ref{hrecursion}). As in our papers \cite{brunelli3,brunelli1,brunelli9} it remains an
interesting question to construct these charges in a more direct
manner.

We can also obtain a Lax representation, in a Gelfand-Dikii form, for the EB hierarchy of equations, in a straightforward way. It is well known (see \cite{brunelli1,camassa} and references therein) that the recursion operator defines a eigenvalue problem of the type
\begin{equation}
(1 - \lambda R^\dagger)\,\phi= 0\;,\label{eigenproblem}
\end{equation}
for the eigenfunction $\phi$ with eigenvalue $1/\lambda$. Since we know the inverse of the recursion operator we obtain from (\ref{eigenproblem})
\begin{equation}
\left[\left(R^\dagger\right)^{-1}-\lambda\right]\,\phi=0\;,\label{ieigenproblem}
\end{equation}
which defines a Lax eigenvalue problem. Using (\ref{invrf}) we identify the Lax operator for our system to be
\[
L\equiv \left(R^\dagger\right)^{-1} \!\!\!\!= \frac{1}{F}\,\partial \frac{1}{F}\, \partial +A_{x}\,\partial^{-1}A_{x}\partial\;.
\]
It can be readily checked that the
hierarchy of EB equations (\ref{localflow}) (up to multiplicative constants) can be obtained from the non-standard
Lax equation
\[
\frac{\partial L}{\partial t_{-n}} = \left[(L^{(2n+1)/2})_{\geq
2},L\right]\;,\quad n=1,2,3,\dots\;.
\]
The conserved quantities (\ref{local}) and $H_0$ in (\ref{nonlocal}) for this system (up to multiplicative constants) can be obtained from
${\rm Tr}\,L^{(2n-1)/2}$, $n=0,1,2,\dots$.

\section{Conclusion:}

In this paper, the Short Pulse and Elastic Beam equations were shown to belong to the same hierarchy corresponding to positive and negative flows. We have shown that these system are bi-Hamiltonian and using recursion we construct infinite series of both local and nonlocal conserved charges as well the respective hierarchy of equations. A Lax pair for the Short Pulse and Elastic Beam equations was derived. The Lax operator yielded the local charges via Adler's trace, however, the construction of the nonlocal charges through this Lax operator is unknown by us. Also, a nonstandard Lax representation in a Gelfand-Dikii form for the Elastic Beam hierarchy of equation was described.

\section*{Acknowledgments}

This work was supported by CNPq (Brazil).


\begin{thebibliography}{99}

\bibitem{schafer}
T. Sch\"afer and C. E. Wayne, {\em Physica D\/} {\bf 196}, 90 (2004).

\bibitem{chung} Y. Chung, C. K. R. T. Jones, T. Sh\"afer and C. E. Wayne, {\em Nonlinearity} {\bf 18}, 1351 (2005).

\bibitem{sakovich} A. Sakovich and S. Sakovich, {\em J. Phys. Soc. Jpn.} {\bf 74}, 239 (2005).

\bibitem{wadati}                                                                         
M. Wadati, K. Konno and Y. H. Ichikawa, {\em J. Phys. Soc. Jpn.} {\bf 47}, 1698 (1979).  

\bibitem{ichikawa}                                                                         
Y. Ichikawa, K. Konno and M. Wadati, {\em J. Phys. Soc. Jpn.} {\bf 50}, 1799 (1981).  

\bibitem{brunelli3}
J. C. Brunelli and G. A. T. F. da Costa, {\em  J. Math. Phys.} {\bf 43}, 6116 (2002).

\bibitem{brunelli1}                                                                  
 J. C. Brunelli, A. Das and Z. Popowicz, {\em  J. Math. Phys.} {\bf 45}, 2646 (2004). 

\bibitem{brunelli9}
J. C. Brunelli and A. Das, {\em  J. Math. Phys.} {\bf 45}, 2633 (2004).

\bibitem{brunelli}
J. C. Brunelli, {\em Comp. Phys. Comm.} {\bf 163}, 22 (2004).

\bibitem{dirac}
P. A. M. Dirac, {\em Lectures on Quantum Mechanics}, Belfer Graduate School of Science Monographs, vol. 2 (New York, 1964); K. Sundermeyer, {\em Constrained Dynamics}, Lecture Notes in Physics, vol. 169 (Springer, Berlin, 1982).

\bibitem{blaszak}
M. B\l aszak, {\em Multi-Hamiltonian Theory of Dynamical Systems} (Springer, Berlin, 1998).

\bibitem{olver}
P. J. Olver, {\em Applications of Lie Groups to Differential Equations},
2nd ed. (Springer, Berlin, 1993).

\bibitem{magri}
F. Magri, {\em J. Math. Phys.\/} {\bf 19}, 1156 (1978).

\bibitem{ghost}
J. A. Sanders and J. P. Wang, {\em Physica\/} {\bf D149}, 1 (2001); J. M. Verosky, {\em J. Math. Phys.\/} {\bf 32}, 1733 (1991); V. A. Andreev and M. V. Shmakova, {\em J. Math. Phys.\/} {\bf 34}, 3491 (1993); S. Y. Lou, {\em J. Math. Phys.\/} {\bf 35}, 2390 (1994).

\bibitem{okubo}
S. Okubo and A. Das, {\em Phys. Lett.} {\bf B209}, 311 (1988); A. Das and S. Okubo, {\em Ann. Phys.} {\bf 190}, 215 (1989).

\bibitem{brunelli2}
J. C. Brunelli and A. Das, {\em Mod. Phys. Lett.\/} {\bf 10A}, 931 (1995).

\bibitem{adler}
M. Adler, {\em Invent. Math.} {\bf 50}, 219 (1979).

\bibitem{camassa}
R. Camassa, D. D. Holm and J. M. Hyman, {\em Adv. Appl. Mech.}
{\bf 31}, 1 (1994).





\end{thebibliography}
\end{document}